\documentclass[aps,preprint]{revtex4}%
\usepackage{amsfonts}
\usepackage{amsmath}
\usepackage{amssymb}
\usepackage{graphicx}%
\begin{document}
\title{Avoiding superluminal propagation of higher spin waves via projectors onto
$W^{2}$ invariant subspaces. }
\author{Mauro Napsuciale}
\affiliation{Insituto de F\'{\i}sica, Univ.\ de Guanajuato, Lomas del Bosque 103,
Fracc.\ Lomas del Campestre, 37150 Leon, Guanajuato, M\'exico}
\author{Mariana Kirchbach}
\affiliation{Instituto de F\'{\i}sica, Univ.\ Aut.\ de San Luis Potos\'{\i}, Av. Manuel
Nava 6, Zona Universitaria, S.L.P.\ 78290, San Luis Potos\'{\i}, M\'exico }

\begin{abstract}
We propose to describe higher spins as invariant subspaces of the Casimir
operators of the Poincar\'{e} Group, $P^{2}$, and the squared Pauli-Lubanski
operator, $W^{2}$, in a properly chosen representation, $\psi
(\mathbf{\mathbf{p}})$ (in momentum space), of the Homogeneous Lorentz Group.
The resulting equation of motion for any field with $s\neq0$ is then just a
specific combination of the respective covariant projectors. We couple
minimally electromagnetism to this equation and show that the corresponding
wave fronts of the classical solutions propagate causally. Furthermore, for
$(s,0)\oplus(0,s)$ representations, the formalism predicts the correct
gyromagnetic factor, $g_{s}=\frac{1}{s}$. The advocated method allows to
describe any higher spin without auxiliary conditions and by one covariant
matrix equation alone. This master equation is only quadratic in the momenta
and its dimensionality is that of $\psi(\mathbf{\mathbf{p}})$. We prove that
the suggested master equation avoids the Velo-Zwanziger problem of
superluminal propagation of higher spin waves and points toward a consistent
description of higher spin quantum fields.

\end{abstract}
\maketitle

\section{Introduction.}

The field theoretical description of interacting particles with spin $>1$ is a
long standing problem. The interaction of a spin $\frac{3}{2}$
Rarita-Schwinger (RS) field minimally coupled to an external electromagnetic
field was shown to be inconsistent more than forty years ago \cite{sudarshan1}%
. Later on, Velo and Zwanziger observed superluminal propagation of the RS
wave front in the presence of a minimally coupled electromagnetic field
\cite{VZ1} and studied also the conditions under which the Proca field
interacting with an external electromagnetic field propagates causally
\cite{VZ2}. After these works many authors have addressed above problem from
different perspectives and for different interactions \cite{todos} and the
general feeling seems to be that it is not possible to construct a consistent
quantum theory for massive particles with $s >$ 1.

At several decades of distance in looking afresh onto the equations of motion
can lead to different understanding of this fundamental problem. Weinberg
emphasizes in his textbook on quantum field theory \cite{Weinberg:mt} that the
equation of motion satisfied by the Dirac field is nothing but the record
about the way how one puts together the two irreducible representations,
(1/2,0), and (0,1/2), of the proper orthochronous Lorentz group to form a
field that transforms invariantly under parity. In a wider understanding, this
means that the equations of motion satisfied by a field are just a consequence
of the properties of the representations of the Homogeneous Lorentz Group
(HLG) chosen by us to accommodate the field and the discrete symmetries we
require to be realized in this space. Closely related arguments can be found,
among others in \cite{WKT}, \cite{ryder}, \cite{MK97}, and \cite{prinind}.

More recently, Refs.~\cite{MK03,Gaby} studied covariant projectors onto
invariant subspaces of the squared Pauli-Lubanski operator in the
representation space of the four-vector--spinor and showed that the associated
equations are free from the Velo-Zwanziger problem. The corresponding
projectors for the $(s,0)\oplus(0,s)$ representation space were studied in
\cite{MC} where it was shown that under minimal coupling a particle in this
representation has the correct value for the spin gyromagnetic factor,
$g_{s}=\frac{1}{s}$, thus proving Belinfante's conjecture \cite{belinfante}
from 1953.

In this work we explore the projectors onto the invariant subspaces of the
Poincar\'{e} Casimir operators, the squared four-momentum and thesquared
Pauli-Lubanski operator, for any $s$, and study propagation of the
corresponding wave fronts along the lines of Refs.~\cite{VZ1,VZ2}. The paper
is organized as follows. In the next Section we recall in brief current
description of higher spins and its relation to the Poincar\'{e} group. In
Section III we suggest to describe higher spins as invariant subspaces of the
Poincar\'{e} Casimirs. In Section IV we show that particles within this
framework propagate causally in the presence of an electromagnetic field, thus
avoiding the classical Velo-Zwanziger problem. The paper closes with a brief Summary.

\section{Current description of fields and its relation to Poincar\'{e} group
representations.}

The primary classification of elementary systems is usually done by
identifying them (up to form factors) with the irreducible representations
(irreps) of the Poincar\'{e} group ($PG$). If so, then one necessarily has to
consider particles as invariant spaces of the Casimir operators of this
group-- the squared four-momentum $P^{2}$, on the one side, and the squared
Pauli-Lubanski operator $W^{2}$, on the other side and label them by their
respective eigenvalues, $p^{2}$, and $-p^{2}s(s+1)$, as $|p^{2},s(s+1)>$.
{}Further quantum numbers can be associated with the Casimir invariants of the
underlying Homogeneous Lorentz Group (HLG), $SO(1,3)$, and are approached by
the reduction chain $PG\supset SO(1,3)$. For finite dimensional
representations, the Casimir invariants of $SO(1,3)$ are frequently expressed
in terms of two $SU(2)$ Casimirs, in turn denoted by $\mbox{\bf S}_{L}^{2}$,
and $\mbox{\bf S}_{R}^{2}$, of $SU(2)_{L}\otimes SU(2)_{R}$, a group that is
locally isomorphic to $SL(2,C)$, the universal covering of HLG. The two
additional quantum labels gained in this manner are the well known left-- and
right handed "angular momenta", $s_{L}$, and $s_{R}$, respectively. Therefore,
a covariant state labeling can be introduced as: $|p^{2},s(s+1);s_{L},s_{R}>$,
with $s=|s_{L}-s_{R}|,...,s_{L}+s_{R}$. In so doing one encounters essentially
two types of finite dimensional HLG representations.

\begin{enumerate}
\item The first ones contain just one $W^{2}$ invariant subspace, and
correspond to the case when one of the $s_{L}$, $s_{R}$ labels vanishes (i.e.
either $(s_{L},0)$, or $(0,s_{R})$), and $s_{R}=s_{L}$. In such a case,
$s_{L/R}(s_{L/R}+1)=s(s+1)$, equals the $\left(  -\frac{1}{m^{2}}W^{2}\right)
$ eigenvalue in the space under consideration (see Eq.~(\ref{W2_rest}) below)
and $W^{2}$ -- and $\mbox{\bf S}_{L/R}^{2}$ invariant spaces coincide. Irreps
of the above type are suggestive of replacing $W^{2}$-- by $SU(2)$ spin labels.

As long as the basic fields in physics are precisely of the above type (the
Dirac field is $(1/2,0)\oplus(0,1/2)$, the electromagnetic field strength
tensor is $(1,0)\oplus(0,1)$, and scalars are just $(0,0)$) identifying
Poincar\'{e} labels with $SU(2)$ spins works out without any harm.

\item The second ones are HLG irreps containing several $W^{2}$ invariant
subspaces. In this case, both $s_{L}$, and $s_{R}$ are non-vanishing, and the
irreps are of the type $\left(  s_{L},s_{R}\right)  $ with $s_{L}\not =0$, and
$s_{R}\not =0$. Examples are the vector--, and tensor gauge fields,
$(1/2,1/2)$, and $(1,1)$, respectively. In the rest frame, $W^{2}=-\frac
{1}{m^{2}}\mbox{ \bf S}^{2}$ hence $W^{2}$ and $\mbox{ \bf S}^{2}$ invariant
sub-spaces coincide. However, beyond rest frame, in flight, $W^{2}$ and
$\mbox{\bf S}^{2}$ invariant sub-spaces are no longer identical, a situation
caused by the property of the boost to mix up SU(2) spins differing by one unit.
\end{enumerate}

Often, Lorentz representations that contain as building blocks irreps of the
second type, appear attractive for the description of higher spins, the
classical examples being the totally symmetric $K$ rank Lorentz tensors with
Dirac spinor components, generically denoted by $\psi_{\mu_{1}...\mu_{K}}$.
They are exploited for the description of fields that have been labeled in the
rest frame by the highest spin $J=K+1/2$. The separation between Lorentz and
spinor indices inherent to such tensors makes them especially appealing for
the construction of covariant fermion-boson vertices. However, one has to face
the problem how to pick up the favored degrees of freedom and exclude
interference with the unwanted ones. It seems inevitable to return back to the
Poincar\'{e} invariants, if one wishes to distinguish all the degrees of
freedom contained in $\psi_{\mu_{1}...\mu_{K}}$ in a covariant and
transitionally invariant fashion. Yet, for one reason or the other, this is
not the path tenaciously pursued by the theory. Rather, one still prefers to
stay within the elaborated scheme of substituting $W^{2}$ by $SU(2)$ labels,
but, yes, modify the latter scheme to account for the new situation in
introducing constraints, considered as appropriate.

To be specific, in order to select out of $\psi_{\mu}$ (a field belonging to
$[(1/2,0)\oplus(0,1/2)]\otimes(1/2,1/2)$) the $W^{2}$ invariant subspace that
relates to spin $3/2$ at rest, one requires
\begin{align}
(i\partial^{\mu}\gamma_{\mu}-m)\psi_{\mu}  &  =0\,,\nonumber\\
\partial^{\mu}\psi_{\mu}  &  =0\,,\nonumber\\
\gamma^{\mu}\psi_{\mu}  &  =0\,. \label{RS_set}%
\end{align}
Exploiting constraints (some times termed to as auxiliary, or, supplementary,
conditions) in place of $W^{2}$ quantum numbers brings the advantage to remain
within the framework of equations linear in the momenta, and to work with
four-dimensional Dirac spinors. However, these advantages reveal themselves as
deceptive at the moment one has to face grave worries about compatibility of
constraints and dynamics. Recall, that the constraints change upon gauging and
one has to make sure that the modification is preserved in time by the
equation of motion and the latter does not violate causality. Notice that
covariance alone is indeed a necessary but not a sufficient condition for
special relativity. For example, space-like intervals are doubtlessly
covariant objects, but they are unacceptable for the description of
\textit{free\/} physical fields as they prescribe the particle to violate
causality during propagation. Precisely a flaw of that very type was revealed
by Velo and Zwanziger in Ref.~\cite{VZ1} regarding the $\gamma^{\mu}\psi_{\mu
}=0$ constraint onto the four--vector spinor. Velo and Zwanziger showed that
above constraint triggers acausal propagation of Rarita-Schwinger particles
crossing an electromagnetic field.

\noindent In the present article we shall avoid above inconsistencies in
developing a different view on form and content of wave equations for higher
spins. Namely, we take the position that the equation of motion for whatever
free particle has to be (i) a function of $P^{2}$ and $W^{2}$, the Casimir
invariants of the Poincar\'{e} group, (ii) operates immediate, i.e. without
any supplementary constraints, on the HLG representation chosen to embed the
field as one of its covariant sectors\footnote{We shall design argumentation
very general and in reference to any one Lorentz representation with more but
one $W^{2}$ invariant subspace. The particular case of $\psi_{\mu_{1}%
,...,\mu_{K}}$ is then automatically accounted for.}.

Within this context, there are two primordial equations of motion to be
satisfied by any field. One of them searches for $P^{2}$ invariant subspaces.
It is nothing more but the Klein-Gordon equation. The other one secures in
addition invariance under pseudo--rotations and pins-down $W^{2}$ invariant
subspaces by means of appropriately constructed covariant projectors. It is
that very latter type of equations on which we focus attention here. {}For the
sake of self-sufficiency of the presentation, the subsequent Section opens
with a brief review of the basics of space-time symmetries.

\section{ Covariant wave equations for higher spins from $W^{2}$ invariant
subspaces.}

\subsection{Basics of Space-Time Transformations.}

A general Poincar\'{e} transformation in space time can be written in the
factorized form
\begin{equation}
g(b,\Lambda)=T(b)\ \Lambda\, ,
\end{equation}
where $T(b)=g(b,E)$ (E denotes the unit matrix) is a translation and
\ $\Lambda=g(0,\Lambda)$ is a proper Lorentz transformation. In the standard
convention, the generators of the translation group in 1+3 time-space
dimensions, $\mathcal{T}$\ $_{1,3}$\ , are $P_{\mu}$ in $T(b)$, which are
commuting,
\begin{equation}
\qquad\left[  P_{\mu},P_{\nu}\right]  =0.
\end{equation}
The HLG transformation in coordinate space,%

\begin{equation}
x_{\mu}^{\prime} =\Lambda_{\mu}^{\quad\nu}\quad x_{\nu},\quad\Lambda_{\mu
}^{\quad\nu} =\exp\left[  -\frac{i}{2}\theta^{\mu\nu}L_{\mu\nu}\right]  \,
,\quad L_{\mu\nu}=X_{\mu}P_{\nu}-X_{\nu}P_{\mu}\, ,
\end{equation}
induces the following transformation for a field $\psi(x)$,%

\begin{equation}
\psi^{\prime}(x)=\exp\left[  -\frac{i}{2}\theta^{\mu\nu}M_{\mu\nu}\right]
\psi(\Lambda^{-1}x).
\end{equation}
Here, $\theta^{\mu\nu}$are continuous parameters, while the $n\times n$
matrices $M_{\mu\nu}$ represent a totally antisymmetric 2nd rank Lorentz
tensor. They are the generators of the homogeneous Lorentz group in the
representation space of interest, and satisfy the commutation relations of the
associated algebra :%
\begin{equation}
\left[  M_{\mu\nu},M_{\alpha\beta}\right]  =-i(g_{\mu\alpha}M_{\nu\beta
}-g_{\mu\beta}M_{\nu\alpha}+g_{\nu\beta}M_{\mu\alpha}-g_{\nu\alpha}M_{\mu
\beta}). \label{M_comm}%
\end{equation}
Their commutators with the generators of the translation group read
\begin{equation}
\left[  P_{\mu},M_{\alpha\beta}\right]  =i(g_{\mu\alpha}P_{\beta}-g_{\mu\beta
}P_{\alpha}), \label{M_P_comm}%
\end{equation}
where $g_{\mu\nu}=diag(1,-1,-1,-1)$ is the metric tensor. The $M_{\mu\nu}$
generators consist of%

\begin{equation}
M_{\mu\nu}=L_{\mu\nu}+S_{\mu\nu},\qquad\left[  L_{\mu\nu},S_{\mu\nu}\right]
=0,
\end{equation}
where $L_{\mu\nu}$, and $S_{\mu\nu}$ in turn generate rotations in external
coordinate-- and internal representation spaces. The generators of boosts
($\mathcal{K}_{x},\mathcal{K}_{y},\mathcal{K}_{z}$) and rotations
($J_{x},J_{y},J_{z}$) are related to $M_{\mu\nu}$ via%

\begin{equation}
\mathcal{K}_{i}=M_{0i}\ ,\qquad J_{i}=\frac{1}{2}\epsilon_{ijk}M_{jk}\ ,
\end{equation}
respectively.

\subsection{Pauli Lubanski Vector and Associated Casimir Invariant.}

The Pauli--Lubanski (PL) vector is now defined as
\begin{equation}
W_{\mu}=-\frac{1}{2}\epsilon_{\mu\nu\alpha\beta}M^{\nu\alpha}P^{\beta},
\label{paulu}%
\end{equation}
where $\epsilon_{0123}=1$. This operator can be shown to satisfy the
commutators
\begin{equation}
\lbrack W_{\alpha},M_{\mu\nu}]=i(g_{\alpha\mu}W_{\nu}-g_{\alpha\nu}W_{\mu
}),\qquad\lbrack W_{\alpha},P_{\mu}]=0, \label{conmrelpl}%
\end{equation}
\noindent i.e. it transforms as a four-vector under Lorentz transformations.
The remarkable point is that the external coordinate part of $M_{\mu\nu}$,
namely the "orbital" momentum \ $L_{\mu\nu}$, does not contribute to $W_{\mu}$
due to the anti-symmetric Levi-Civita tensor. As a result, $W_{\mu}$ restricts
to
\begin{equation}
W_{\mu}=-{\frac{1}{2}}\epsilon_{\mu\nu\rho\tau}S^{\nu\rho}P^{\tau}\,,
\label{PauLu1}%
\end{equation}
and its squared (in covariant form) is calculated to be
\begin{equation}
W^{2}=-{\frac{1}{2}}S_{\mu\nu}S^{\mu\nu}P^{2}+G^{2}\,,\quad G_{\mu}:=S_{\mu
\nu}P^{\nu}\,. \label{paulu2}%
\end{equation}
The operators $S_{\mu\nu}$ act exclusively in the internal spin space and
commute like
\begin{equation}
\left[  S_{\mu\nu},S_{\alpha\beta}\right]  =-i(g_{\mu\alpha}S_{\nu\beta
}-g_{\mu\beta}S_{\nu\alpha}+g_{\nu\beta}S_{\mu\alpha}-g_{\nu\alpha}S_{\mu
\beta})\,. \label{S_comm}%
\end{equation}
As long as Eq.~(\ref{S_comm}) has same form as Eq.~(\ref{M_comm}), one may
view $S_{\mu\nu}$ as generators of Lorentz transformations in the intrinsic
space. However, in contrast to Eq.~(\ref{M_P_comm}), $S_{\mu\nu}$
\textit{commute\/} with the operators of translations
\begin{equation}
\quad\left[  P_{\alpha},S_{\mu\nu}\right]  =0\,. \label{Poinc_contr}%
\end{equation}
In effect, one does not find precisely Poincar\'{e} transformations in the
internal space but rather a contracted form of them. Hereafter we will refer
to the group generated by $S_{\mu\nu}$ as the \textquotedblleft Internal
Homogeneous Lorentz Group\textquotedblright\ ($\mathcal{I}$HLG) to distinguish
it from the HLG spanned by $M_{\mu\nu}$. In summary, one can write down
generators of boosts and rotations in the internal space as
\begin{equation}
K_{i}=S_{0i},\qquad\ S_{i}=\frac{1}{2}\epsilon_{ijk}S_{jk}\,.
\end{equation}
The internal HLG has by itself two Casimir invariants, in turn given by
$C_{1}={\frac{1}{4}}S_{\mu\nu}S^{\mu\nu}$, and $C_{2}=$ $S_{\mu\nu}%
\widetilde{S}^{\mu\nu}$, with $\widetilde{S}_{\mu\nu}=\epsilon_{\mu\nu\rho
\tau}S^{\rho\tau}$. In terms of $\mbox{\bf K}$, and $\mbox{\bf S}$ one finds
\begin{equation}
C_{1}=\frac{1}{2}(\mathbf{S}^{2}-\mathbf{K}^{2})\,,\quad C_{2}=i\mathbf{S\cdot
K}\,. \label{casimirs}%
\end{equation}
The latter equation allows to cast $W^{2}$ into the form
\begin{equation}
W^{2}=-2C_{1}P^{2}+G^{2}\ . \label{paso_1}%
\end{equation}
{}For irreps of the type $(s,0)\oplus(0,s)$ where $K_{i}=\mp iS_{i}$, one
finds the insightful relation \cite{Gaby}
\begin{equation}
G^{2}=-W^{2}\,. \label{P:_s0_0s}%
\end{equation}
Insertion of Eq.~(\ref{P:_s0_0s}) into Eq.~(\ref{paso_1}) amounts to
\begin{equation}
W^{2}=-\mathbf{S}^{2}P^{2}\,. \label{W2_rest}%
\end{equation}
The latter relation explains the privileged position of $(s,0)\oplus(0,s)$
states to carry unique SU(2) spin both at rest (where $W^{2}$ any way reduces
to $-\mathbf{\mathbf{S}}^{2}\,m^{2}$ in accord with Eq.~(\ref{W2_rest})) and
in flight. However, for all the other types of Lorentz representations,
$W^{2}\not =G^{2}$ and the $\left(  -\frac{1}{m^{2}}W^{2}\right)  $ labels for
particles in flight do not have the interpretation of ordinary $SU(2)$ spin.
In the following we label $W^{2}$ invariant sub-spaces by $s$ but in general
without any reference to SU(2) spin.

\subsection{Covariant projectors onto $W^{2}$ invariant subspaces.}

To begin with we recall that the interpretation of elementary systems as
Poincar\'{e} group irreducible representations requires any field to transform
invariantly under the action of \ both $P^{2}$ and $W^{2}$. In the following
we work with massive fields. The former invariance leads to the Klein-Gordon
equation for any arbitrary field
\begin{equation}
\left(  P^{2}-m^{2}\right)  \psi(\mathbf{\mathbf{p}})=0\,. \label{kl-gr}%
\end{equation}
Invariance under the action of $W^{2}$ results into the new condition%
\begin{equation}
\Pi^{s}(\mathbf{\mathbf{p}})\psi(\mathbf{\mathbf{p}})=\psi(\mathbf{\mathbf{p}%
})\,, \label{dummy}%
\end{equation}
where $\Pi^{s}(\mathbf{\mathbf{p}})$ stands for an appropriately constructed
covariant projector onto the $\left(  -p^{2}s(s+1)\right)  $ invariant
subspace of $W^{2}$ in $\psi(\mathbf{\mathbf{p}})$. To be specific, for the
case of the four-vector spinor, such projectors have been presented in
Ref.~\cite{MK03}. In general, equations of the type (\ref{dummy}) are
equivalent to
\begin{equation}
\left[  W^{2}+P^{2}s(s+1)\right]  \ \psi(\mathbf{\mathbf{p}})=0.
\label{master}%
\end{equation}
Next, it is necessary to account for the mass shell condition in
Eq.~(\ref{kl-gr}). For this purpose, we sum up Eqs.~(\ref{master}) and
(\ref{kl-gr}) to obtain%
\begin{equation}
\left[  \frac{1}{s}W^{2}+sP^{2}+m^{2}\right]  \ \psi(\mathbf{\mathbf{p}})=0,
\label{master1}%
\end{equation}
and cast the latter equation into the explicitly covariant form%
\begin{equation}
\left[  t_{\mu\nu}P^{\mu}P^{\nu}-m^{2}\right]  \psi(\mathbf{\mathbf{p}})=0\,.
\label{ma-eq}%
\end{equation}
Here $t_{\mu\nu}$ stands for%
\[
t_{\mu\nu}=\frac{1}{s}(2C_{1}g_{\mu\nu}-S_{\alpha\nu}S^{\alpha}\,_{\mu
})\ -s\ g_{\mu\nu}\,,
\]
$C_{1}$ denotes the first Casimir in Eq.(\ref{casimirs}) and $S^{\beta\rho}$
are the $\mathcal{I}HLG$ generators in the particular representation chosen
for $\psi(\mathbf{\mathbf{p}})$. Their construction as solutions of the
algebra of the Lorentz group for the representation space under consideration
is straightforward \cite{ryder,prinind,MK03}.

Using now the gauge principle for electromagnetism in this equation we obtain%
\begin{equation}
\left[  \left(  \frac{1}{s}(2C_{1}g_{\mu\nu}-S_{\alpha\nu}S^{\alpha}\,_{\mu
})-s\ g_{\mu\nu}\right)  \ \pi^{\mu}\ \pi^{\nu}-m^{2}\right]  \psi
(\mathbf{\mathbf{p}})=0, \label{eom}%
\end{equation}
with $\pi^{\mu}=P^{\mu}+eA^{\mu}$, and $e$ denoting the charge of the field
$\psi(\mathbf{\mathbf{p}})$. Notice that Eq.~(\ref{eom}) is a covariant matrix
equation that operates in the vector space of the dimensionality of
$\psi(\mathbf{\mathbf{p}})$. For example, when $\psi(\mathbf{\mathbf{p}})$
stands for the four-vector-- spinor, $W^{2}$ is represented by a $16\times16$
matrix. \ For the sake of illustration, we here bring the Lagrangian density
\ for the lowest Rarita-Schwinger representation. It reads%

\begin{equation}
\mathcal{L}(x)=\overline{\psi}(x)\ t_{\mu\nu}\ \pi^{\mu}\ \pi^{\nu}%
\ \psi(x)-m^{2}\overline{\psi}(x)\psi(x),\label{Lagr_RS}%
\end{equation}
where $\overline{\psi}(x)=\psi^{\dagger}(x)(\gamma^{0}\otimes g)$ where $g$ is
the matrix of the metric tensor.\ The definition of $\overline{\psi}(x)$ has
to be performed for each representation individually. When applied to the
Dirac representation $(\frac{1}{2},0)\oplus(0,\frac{1}{2})$ , Eq. (\ref{eom})
also has the great advantage to yield the correct value of the gyromagnetic
factor, $g_{s}=2$. This is not fortuitous but reflects the general property of
our master equation (\ref{eom}) to predict the correct value for the
gyromagnetic ratio as $g_{s}=\frac{1}{s}$ for fields in $(s,0)\oplus(0,s)$
\cite{MC}. Had we used instead Eq.~(\ref{master}) alone, we would have found
the problematic case of $g_{s}=\frac{1}{s(s+1)}$.

With respect to $(\frac{1}{2},0)\oplus(0,\frac{1}{2})$, Eq.(\ref{master1}) is
nothing more but the Klein-Gordon equation for each field component. This is
due to the fact that the squared Pauli-Lubanski vector for all $(s,0)\oplus
(0,s)$ fields is just $-s(s+1)P^{2}\mathbf{1}_{(2s+1)\times(2s+1)}$. The
$W^{2}$ Casimir invariant identifies only the spin content and remains
indifferent to the discrete $C$, $P$, or, $T$ properties of the representation
of interest. Recall that one has different options to stick together, say,
$(\frac{1}{2},0)$ and $(0,\frac{1}{2})$ in depending on whether one wants
$(\frac{1}{2},0)\oplus(0,\frac{1}{2})$ to diagonalize the parity--,
$\gamma^{0}\mathcal{R}$ or, the charge conjugation--, $i\gamma_{2}K$,
operator. For parity eigenstates one ends up with the standard Dirac equation,
while for $C-$ parity states one finds again the Dirac equation but with a
Majorana mass term, respectively. Notice however that, under gauging, this
equation gives the right magnetic properties for $(s,0)\oplus(0,s)$ fields .
This means that solutions to Dirac equation are solutions to Eq.(\ref{eom})
although the converse is not necessarily true since our equation specifies
just the value of the spin.

For product representation spaces of the type $\psi_{\mu_{1}\mu_{2}...\mu_{K}%
}$, the most interesting representation space for applications in hadron
physics, the situation is different provided, one is tracking the highest
spin. As long as the highest spins are non-degenerate, there is no confusion
with parity doubling, as would be the case for the lower spins. For these
reasons, Eq.~(\ref{eom}) has its major merits with respect to the highest
spins in the representations.

\noindent Next we study wave front propagation of particles described by means
of Eq.~(\ref{eom}) along the line of Refs.~\cite{VZ1,VZ2}.

\section{Avoiding superluminal propagation of higher spin waves.}

Wave propagation is associated with a hyperbolic system of partial
differential equations \cite{CH}. For such a class of differential equations
the initial value problem can be posed on a class of surfaces ( "space like"
surfaces with respect to the equation of motion). The equations possess
solutions with wave fronts traveling along rays at finite velocities. At any
point on the surface, the rays form a cone that is entirely determined by the
coefficients of the highest derivatives in the equation of motion \cite{CH}.
The wave front can be characterized by $n_{\text{ }}^{\mu}=(n^{0},\mathbf{n)}%
$, the vectors normal to the characteristic surface. The system of equations
is hyperbolic if $n^{0}$ is real for any $\mathbf{n}$. To find the normal
vectors it is sufficient to first replace in the highest derivatives of the
equation of motion $P_{\mu}$ by $n_{\mu\text{ }}$ and then calculate the
determinant $D(n)$ (so called "characteristic determinant" \cite{VZ2}) of the
matrix given by the corresponding coefficients.

\subsection{Wave front propagation of the Klein-Gordon, Dirac and
Rarita-Schwinger equations.}

In cases when the coupling to external fields is carried by the lower
derivatives in the equation of motion, such as, say, the Klein--Gordon
equation, the ray cones for interacting and free fields coincide. Indeed, in
the latter case and under minimal coupling one finds
\begin{equation}
\left[  \pi^{\mu}\pi_{\mu}-m^{2}\right]  \psi(\mathbf{p})=\left[  P^{\mu
}P_{\mu}+e(P^{\mu}A_{\mu}+A^{\mu}P_{\mu})+e^{2}A^{\mu}A_{\mu}-m^{2}\right]
\psi(\mathbf{p})=0. \label{KL_Go}%
\end{equation}
The vanishing of the characteristic determinant in this case yields%
\begin{equation}
D(n)=\mbox{Det}(n^{2})=n^{2}=0\,, \label{DET_KLG}%
\end{equation}
which has real $n^{0}$ for any $\mathbf{n}$. Same is true for Dirac particles,
though not as obvious. As is well known, a Dirac particle coupled minimally to
the electromagnetic field is described by
\begin{equation}
\left[  \gamma^{\mu}(P_{\mu}+eA_{\mu})-m\right]  \psi(\mathbf{\mathbf{p}%
})=0\,. \label{DIR_EQ}%
\end{equation}
Now, the resulting characteristic determinant is found to be the squared of
Eq.~(\ref{DET_KLG})
\begin{equation}
D(n)=\mbox{Det}(\gamma^{\mu}n_{\mu})=(n^{2})^{2}\,. \label{DET_DIR}%
\end{equation}
The vanishing of this determinant results once again into a ray cone that
coincides with the light cone.

\noindent The wave front propagation of the solution of the Rarita-Schwinger
set of equations was studied in great detail in Ref.~\cite{VZ1}. To understand
the essence of the latter work recall that Eqs. (\ref{RS_set}) or the
analogous equation in the interacting case, can be derived from a Lagrangian,
a method suggested by Fierz and Pauli \cite{FP}. Within the latter framework
not all the Euler-Lagrange equations appear as genuine equations of motion,
meaning that some of them may not involve time derivatives, a property that
qualifies them only as constraints onto the fields. Precisely this is the case
for the Rarita-Schwinger framework discussed in Section II. As a consequence,
any surface in space-time is a characteristic surface \cite{CH}. The
Rarita-Schwinger system of coupled equations turns to be equivalent to a
system of hyperbolic equations supplemented by constraints that are conserved
in time. In this case, the wave fronts of the constrained system are no longer
given by the characteristic determinant of the Euler-Lagrange equations.
Rather, it is necessary to find the genuine equation of motion, i.e. the one
which (i) contains all the higher order derivatives needed for the complete
characterization of the system, (ii) preserves the constraints in time.
Finding such an equation in general introduces, in addition to the derivatives
already present in the system of coupled equations, also new ones which as a
rule spoil causal propagation, a result due to \cite{VZ1,VZ2}.

\subsection{Wave front propagation of $W^{2}$ invariant subspaces.}

In the present work we suggested an alternative formalism to the
Rarita-Schwinger framework. Our proposal was to pin down the degrees of
freedom of interest by means of Eq.~(\ref{eom}). This equation was build upon
the covariant projector onto the $W^{2}$ invariant vector spaces in the
representation under consideration, and did not invoke any supplementary
conditions. In this concern, it is worth to remark that the formalism does not
deal with the whole representation space but only with one of its $W^{2}$
invariant sub-spaces. Below we prove that equations of the latter type do not
suffer the Velo-Zwanziger problem upon gauging.

Firstly, we have to check that for all the degrees of freedom of
$\psi(\mathbf{p})$, the second order time-derivatives enter Eq.~(\ref{eom})
with non-vanishing coefficients. This can be done in full generality in
momentum space where
\begin{equation}
t_{00}=\frac{1}{s}(2C_{1}g_{00}-S_{\alpha0}S^{\alpha}\,_{0})-sg_{00}%
=\mathbf{1}. \label{t00}%
\end{equation}
Therefore, for all $W^{2}$ invariant subspaces, the time derivative of each
field component in Eq.~(\ref{eom}) does not vanish. This equation will be
hyperbolic if the solutions $n^{0}$ to $D(n)=0$ are real for any $\mathbf{n}$.
In this case we must solve
\begin{equation}
\mbox{Det}\left[  -\frac{1}{s}W^{2}(n)-s\ n^{2}\right]  =0.
\label{Determinant}%
\end{equation}
In order to demonstrate that (\ref{eom}) is a hyperbolic equation in the HLG
representation space chosen for $\psi(\mathbf{p})$ we here exploit
decomposition of the latter into invariant subspaces of $W^{2}$.

The most transparent representation of $W^{2}$ is obtained in the basis of
$\mathbf{p}$-dependent $W^{2}$ eigenstates where $W^{2}$ is block diagonal and
equal to
\begin{equation}
W^{2}(P)=-P^{2}\mbox{Diag}\left[  s_{1}(s_{1}+1)\mathbf{1}_{s_{1}}%
,\ s_{2}(s_{2}+1)\mathbf{1}_{s_{2}},...s_{N}(s_{N}+1)\mathbf{1}_{s_{N}%
}\right]  \,. \label{sis}%
\end{equation}
Here $\left\{  s_{1},s_{2}...s_{N}\right\}  $ label the different
eigensubspaces of $W^{2}$ (one of them being $s$) in the representation of
interest, while $\mathbf{1}_{s_{j}}$ denotes the unit matrix of dimensionality
$(2s_{j}+1)\times(2s_{j}+1)$. Notice that the dimensionality, $(d)$, of the
representation space $\psi(\mathbf{p})$ relates to the $W^{2}$ quantum numbers
via $d=\sum_{i}m_{i}(2s_{i}+1)$, where $m_{i}$ is the multiplicity of $s_{i}$.
The latter accounts for possible degeneracies of the $W^{2}$ invariant
subspaces in $\psi(\mathbf{p})$ with respect to further symmetries such like,
say, one of the discrete space--time symmetries.

The determinant (\ref{Determinant}) is calculated as%
\begin{equation}
\mbox{Det}\left[  -\frac{1}{s}W^{2}(n)-s\ n^{2}\right]  =%
%TCIMACRO{\dprod \limits_{k=0}^{N}}%
%BeginExpansion
{\displaystyle\prod\limits_{k=0}^{N}}
%EndExpansion
\left(  n^{2}(\frac{s_{k}(s_{k}+1)}{s}-s)\right)  ^{2s_{k}+1}. \label{Det2}%
\end{equation}

As long as for the integer/half-integer $s$ under consideration, there are no
positive integers and half-integers $s_{k}$ satisfying%
\begin{equation}
\frac{s_{k}(s_{k}+1)}{s}-s=0,
\end{equation}
the roots of the characteristic determinant are $n^{2}=0$. Thus the solutions
have $n^{0}$ real for any $\mathbf{n}$, and Eq.(\ref{eom}) is a set of
hyperbolic equations for the $\psi(\mathbf{p})$ components. The characteristic
surfaces are same for free and interacting particles, and the ray cone
coincides with the light cone. In other words, the wave front propagation of
$W^{2}$ invariant subspaces is free from the Velo--Zwanziger problem.

\section{Conclusions and perspectives.}

In the present article we advocate the idea to consider higher spins as
invariant subspaces of the Casimir operators of the Poincar\'{e} group, the
squared four-momentum and the squared Pauli-Lubanski vector, in a properly
chosen representation of the HLG, $\psi(\mathbf{p})$. In executing the idea we
demonstrated that any higher spin is described in terms of one covariant
matrix equation that (i) is determined exclusively by the HLG generators in
$\psi(\mathbf{p})$, (ii) is of the dimensionality of $\psi(\mathbf{p})$, (iii)
is always of second order in the momenta. We gauged this equation minimally
and found the resulting particle propagation to be causal, thus avoiding the
classical Velo-Zwanziger problem. Moreover, for the single spin valued
$(s,0)\oplus(0,s)$ representations, our master equation (\ref{eom}) has the
great advantage to predict the correct value for the gyromagnetic ratio,
$g_{s}=\frac{1}{s}$, thus proving Belinfante's conjecture \cite{belinfante}
from 1953.

The development of a calculation scheme for interacting particles of higher
spins from the perspective of the present work is underway.

\section{Acknowledgments}

Work supported by Consejo Nacional de Ciencia y Tecnologia (CONACYT) Mexico
under projects 37234-E and C01-39820 .

\end{document}